# Research on mouse brain tissue optical model based on Monte Carlo

Xianlin Song [a, #, *], Rui Wang [b, #], Jianshuang Wei [c, d, #], Lingfang Song [e]

[a] School of Information Engineering, Nanchang University, Nanchang 330031, China;
[b] Ji luan Academy, Nanchang University, Nanchang 330031, China;
[c] Britton Chance Center for Biomedical Photonics, Wuhan National Laboratory for Optoelectronics-Huazhong University of Science and Technology, Wuhan 430074, China;
[d] Moe Key Laboratory of Biomedical Photonics of Ministry of Education, Department of Biomedical Engineering, Huazhong University of Science and Technology, Wuhan 430074, China;
[e] Nanchang Normal University, Nanchang 330031, China;
[#] equally contributed to this work
* Corresponding author: songxianlin@ncu.edu.cn

**ABSTRACT**

The brain is composed of the cerebrum, cerebellum, diencephalon and brainstem. The cerebrum is the superlative part of the central nervous system and also the main part of the brain. There are differences and similarities between humans and mouse. The study of mouse brain model is helpful to understand the process in clinical trials and also has reference significance for the study of human brain. Therefore, the study of mouse brain is particularly important. As the skull has a large scattering effect on light, it is difficult for us to image the brain through the skull directly. Therefore, we often use methods such as optical clearing or thin skull to reduce or remove the influence of the skull on imaging. In this paper, the transmission of photons in mouse brain was studied using Monte Carlo method. In the study of photon trajectories, the photon distribution without intact skull went farther in both longitudinal and transverse directions compared with that of with intact skull. In terms of the optical absorption density and fluence rate. On the condition of with intact skull, the distribution of optical absorption density and fluence rate was fusiform and rounder on the whole. The radial distribution range of optical absorption density and fluence rate was 0.25 cm, which was approximately 2.5 times of that of with intact skull. In the depth direction, due to the strong scattering and absorption of the scalp and skull, the optical absorption density dropped sharply from 0.890 $cm^{-1}$ to 0.415 $cm^{-1}$. When the photons arrived at the gray matter layer, only a few photons were reserved. Due to the strong absorption and scattering effect of the gray matter layer, only a few photons left, the optical absorption density increased from 0.415 $cm^{-1}$ to 0.592 $cm^{-1}$, and then decreased again. When the depth was 1.35 cm, the optical absorption density dropped to 0 $cm^{-1}$. After removing the skull, due to the weak absorption and scattering effect of normal saline and cerebrospinal fluid, the optical absorption density was low (0.119 $cm^{-1}$) and dropped slowly. When the photons arrived at the gray matter layer, most of the photons were reserved. Due to the strong absorption and scattering effect of the gray matter layer, the optical absorption density increased from 0.117 $cm^{-1}$ to 0.812 $cm^{-1}$, then the optical absorption density decreased to 0 $cm^{-1}$ at a depth of 1.35 cm. The distribution of radiant fluence rate is similar to that of optical absorption density. This study will provide reference and theoretical guidance for the optical imaging of mouse brain and the study of the mouse and human brain.

## 1. INTRODUCTION

The brain is the main part of the central nervous system, including the brain, cerebellum, brain stem and other structures [1]. The structure of the brain is intricate and complex, with hundreds of billions of neurons. There are many nerve centers composed of nerve cells, and a large number of up and down nerve fiber bundles pass through and connect the brain, cerebellum and spinal cord. The study of mouse brain model is helpful to understand the process in clinical trials, and also has reference significance for the study of human brain.

There are many imaging techniques for the brain clinically. X-ray computed tomography (X-CT) [2,3], which is commonly used clinically, requires additional contrast agents to visualize blood vessels, and has ionizing radiation due to the use of X-rays. Ultrasound imaging (US) [4] is widely used in clinical testing due to its non-radiation to the human body and flexible movement. However, its resolution is poor (500-1000 $\mu m$), contrast is low, and it cannot Obtain blood oxygen saturation information. Positron emission tomography (PET) [5] is more expensive and has low spatial resolution (at the 1 mm level), and it needs to be combined with CT and other technologies to provide structure information. Magnetic resonance imaging (MRI) [6] is also expensive to use, and the spatial resolution is not high, about 1 mm.

Compared with the above-mentioned clinical imaging technology, optical imaging technology has the characteristics of high resolution. Commonly used optical imaging methods in clinical practice include optical coherence tomography (OCT) [7], intravital microscopy (IVM), confocal laser scanning microscope (CLSM) [8] and two-photon microscopy [9] (TPM) and so on. However, because the skull has a strong scattering effect on light, it is difficult for photons to penetrate the skull to reach the depths, so the use of these optical imaging modes is also limited. When imaging the brain, methods such as thinning the skull, optical clearing, and removing the skull are often used to directly image the structure and function of the brain.

In this article, in order to explore the impact of the skull on the transmission of photons, we use the Monte Carlo method to simulate the movement of photons in mouse brain. We have studied the propagation of photons in mouse brain tissue with and without intact skull, respectively, and the corresponding optical absorption distribution, fluence rate, and photon track were also analyzed. This study will provide reference and theoretical guidance for the optical imaging of mouse brain and the study of the differences and connections between mouse brain and human brain.

## 2. METHODS

### 2.1 Introduction to characteristics of brain tissue

Brain tissue is a multi-layered medium, and the structure and function of each layer are different, and the optical properties are also different. The mouse brain is similar to the human brain in structure. When light is irradiated to the brain from the outside, photons need to pass through the scalp, skull, cerebrospinal fluid, brain and other media to reach the depths of the brain, as shown in Figure 1. In the Monte Carlo simulation of the brain, we use a semi-infinite model with infinity in the horizontal direction and finite depth in the vertical direction to simulate the transmission of photons in the brain tissue. The optical parameters with skull are shown in Table 1.

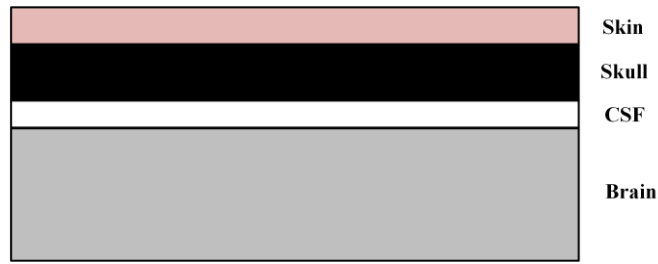

Figure 1. Multi-layer media model of mouse brain.

Table 1. Optical parameters and thickness of four layers of mouse brain tissue with intact skull

| Tissue Type | d(cm) | $\mu_a$ (cm$^{-1}$) | $\mu_s$ (cm$^{-1}$) | g | n |
|---|---|---|---|---|---|
| Skin | 0.1 | 0.191 | 66 | 0.9 | 1.37 |
| Skull | 0.1 | 0.136 | 86 | 0.9 | 1.37 |
| CSF | 0.05 | 0.026 | 0.1 | 0.9 | 1.37 |
| Brain | 1 | 0.186 | 111 | 0.9 | 1.37 |

It can be seen that the brain contains a total of four layers of tissues, and their anisotropy factors are all 0.9, their refractive index are all 1.37. The first layer is the skin. Its depth is 0.1 cm, absorption coefficient is 0.191 cm$^{-1}$, a scattering coefficient is 66 cm$^{-1}$. The second layer is the skull. Its depth is 0.1 cm, absorption coefficient is 0.136 cm$^{-1}$, scattering coefficient is 86 cm$^{-1}$. The third layer is CSF. Its depth is 0.05 cm. The absorption coefficient is 0.026 cm$^{-1}$, the scattering coefficient is 0.1 cm$^{-1}$. The fourth layer is brain. Its depth is 1 cm, absorption coefficient is 0.186 cm$^{-1}$, scattering coefficient is 111 cm$^{-1}$.

After removing the scalp and skull, we covered the head with saline to maintain the brain microenvironment. The optical parameters without intact skull are shown in Table 2.

Table 2 Optical parameters and thickness of four layers of mouse brain tissue without intact skull

| Tissue Type | d(cm) | $\mu_a$ (cm$^{-1}$) | $\mu_s$ (cm$^{-1}$) | g | n |
|---|---|---|---|---|---|
| Normal saline | 0.1 | 0.026 | 8 | 0.9 | 1.37 |
| Normal saline | 0.1 | 0.026 | 8 | 0.9 | 1.37 |
| CSF | 0.05 | 0.026 | 8 | 0.9 | 1.37 |
| Brain | 1 | 0.186 | 111 | 0.9 | 1.37 |

## 2.2 Monte Carlo method

The Monte Carlo method is a statistical simulation method [10]. It is also known as the random sampling or statistical test method. It is a method of using repeated statistical experiments to solve physical and mathematical problems. Such problems can be described directly or indirectly by a random process. The Monte Carl method is used to simulate the transmission process of photons in biological tissues, and has become the main method for studying light transmission in biological tissues.

The basic idea of Monte Carl method is as follows. The principle of photon absorption and scattering is used to track the process of photon passing through chaotic medium. Then, the distribution of energy inside the tissue is obtained by counting the simulation results of a large number of photons.

We consider a random variable $x$, which may be the random step taken by the photon between two successive interactions of the photon in the biological tissue, or the angle at which a photon is deflected due to the occurrence of a scattering event. The probability density function defines the distribution of the variable $x$ over the interval (a, b). The normalized probability density function is:

$$\int_b^a p(x)dx = 1 \tag{1}$$

The computer provides a random variable $\xi$ uniformly distributed in the interval (0, 1), the cumulative distribution function of the variable $\xi$ is:

$$F_\xi(\xi) = \begin{cases} 0 & if \quad \xi \leq 0 \\ \xi & if \quad 0 \leq \xi \leq 1 \\ 1 & if \quad \xi \geq 1 \end{cases} \tag{2}$$

First, step size simulation is carried out, where $u_a$ is the absorption coefficient and $u_s$ is scattering coefficient, then $u_t = u_a + u_s$, the formula of step size $s$ can be obtained:

$$s = \frac{-\ln(1-\xi)}{u_t} = \frac{-\ln(\xi)}{u_t} \tag{3}$$

In a multilayer tissue, the photon may experience multiple motions between layers within the tissue. At this moment: $\sum_i u_{t_i} s_i = -\ln(\xi)$. The deviation angle $\theta$, direction cosine $u$ and azimuth $\psi$ were also simulated, then we get:

$$u = \frac{1}{2g}[1 + g^2 - (\frac{1-g^2}{1-g+2g\xi})^2](g \neq 0) \tag{4}$$

When $g = 0$:

$$u = 2\xi - 1 \tag{5}$$

Convert to scattering deflection angle θ, there are:

$$\theta = \begin{cases} \cos^{-1}\left(\dfrac{1}{2g}\left[1+g^2 - \left(\dfrac{1-g^2}{1+g+2g\xi}\right)^2\right]\right) & (g \neq 0) \\ \cos^{-1}(2\xi - 1) & (g = 0) \end{cases} \tag{6}$$

Since the scattering has axisymmetric property, then: $\varphi = 2\pi\xi$.

First, there is a photon, its weight w is 1, the initial position is (0,0,0), and the initial direction cosine is (0,0,1). The refractive index of the external medium and the surface of the biological tissue are $n_0$ and $n_1$ respectively, then the emission coefficient $R_{sp}$ of specular reflection is $R_{sp} = \dfrac{(n_0 + n_1)^2}{(n_0 + n_1)^2}$. Supposing the current position of the incident photon is (x, y, z), and the current propagation direction is determined by its direction cosine $(u_x, u_y, u_z)$. Then the next position of the photon is determined by the following formula:

$$\begin{cases} x' = x + u_x s \\ y' = y + u_y s \\ z' = z + u_z s \end{cases} \tag{7}$$

Supposing the initial weight of the photon is $w_0$, and the weight after the nth interaction is $w_n$, then $w_n = w_0\left(\dfrac{u_s}{u_t}\right)^n$, the new direction cosine of the photon is given as follows:

$$\begin{cases} u_x' = \dfrac{\sin\theta}{\sqrt{1-u_z^2}}(u_x u_z \cos\varphi - u_y \sin\varphi) + u_x \cos\theta \\ u_y' = \dfrac{\sin\theta}{\sqrt{1-u_z^2}}(u_y u_z \cos\varphi - u_x \sin\varphi) + u_y \cos\theta \\ u_z' = -\sin\theta \cos\varphi\sqrt{1-u_z^2} + u_z \cos\theta \end{cases} \tag{8}$$

If the motion direction of the photon is very close to the normal direction of the biological tissue surface, the new direction cosine of the photon is:

$$\begin{cases} u_x' = \sin\theta \cos\varphi \\ u_y' = \sin\theta \cos\varphi \\ u_z' = SIGH(u_z)\cos\theta \end{cases} \tag{9}$$

When the photon reaches the boundary of the biological tissue, the distance from the current position (x, y, z) to the boundary of the biological tissue along its propagation direction is defined as the reduction step $s_1$. $z_0$ and $z_1$ is the z coordinate of the upper and lower boundaries of the organization in the cartesian coordinate system, and z is the z coordinate of the scattered point of the photon before it reaches the boundary:

$$\begin{cases} s_1 = \dfrac{(z_0 - z)}{u_z} & u_z \leq 0 \\ s_1 = \dfrac{(z_1 - z)}{u_z} & u_z > 0 \end{cases} \tag{10}$$

Criteria for whether the photon has reached the tissue boundary:

1. When $s < s_1$, the photons do not reach the organizational boundary, which will continue to spread in the organization.

2. When $s > s_1$, the photons reach the organizational boundary.

If the incident angle of the photon is $\alpha_i$ and the projection angle is $\alpha_t$, then $\alpha_i = \cos^{-1}(|u_z|)$, $n_i \sin\alpha_i = n_t \sin\alpha_t$, the reflectivity of the total reflection of the photon arriving at the boundary can be obtained by Fresnel formula:

$$R(\alpha_i) = \frac{1}{2}[\frac{\sin^2(\alpha_i - \alpha_t)}{\sin^2(\alpha_i + \alpha_t)} + \frac{\tan^2(\alpha_i - \alpha_t)}{\tan^2(\alpha_i - \alpha_t)}] \qquad (11)$$

When $\xi \leq R(\alpha_i)$, total internal reflection of photons occurs. When $\xi > R(\alpha_i)$, photons escape from tissue.

After total reflection, the direction cosine is updated to $(u_x, u_y, -u_z)$, at which time the remaining step size is $s - s_1$. If it is large enough, it can reach other boundaries and repeat the above process. If not, the photons travel through the tissues and are absorbed and scattered.

We use the roulette method to determine the termination of the photon. First, generate an integer m, and then generate a uniformly distributed random number $\xi$. If $\xi \leq 1/m$, the weight of the photon is updated to $mw$, the photon will continue to transmit with the new weight; If $\xi > 1/m$, the weight of the photon is 0, the photon tracking is ended. For multilayer biological tissues, transmission and reflection occur at the interface of each layer, $(u_{\alpha_n}, u_{s_n}, g_n)$ representing absorption coefficient, reflection coefficient and individuality factor of the nth layer, respectively.

### 2.3 Steps of simulation experiment

In this paper, the Monte Carlo simulation program prepared by Dr. Lihong Wang is used to configure the environment, including the configuration of photon number, the lattice spacing *dr* and *dz* of the two-dimensional grid system, the number of grids, the number of tissue layers and the optical parameters of each layer. In this simulation process, the number of photons was set as 10000, the lattice spacing *dr* and *dz* of the two-dimensional grid system were set as 0.1 cm and 0.1 cm, the number of grids was set as 400 and 200, and the number of tissue layers was set as 4. The optical parameters of with intact skull and without intact skull were substituted in the model, and finally, the parameters such as fluence rate and optical absorption density were obtained.

## 3. RESULTS

### 3.1 The path of photon propagation in the brain

Biological tissue is a high scattering random medium, and the propagation of photons in different tissues varies greatly. Figure 2(a) and Figure 2(b) simulate the photon track in the brain tissue without intact skull and with intact skull, respectively. It can be seen from the figure that the distribution of photons in the z direction in Figure 2 (a) is deeper than that in Figure 2(b), so the depth of photon movement in the brain is larger. Since the skin and skull have a large scattering and absorption effect on photons, and the CSF layer is a medium with low absorption and low scattering effect, so the photons can move to deeper tissues without intact skull. In Figure 2(a), the lateral range of photon motion is larger than that in Figure 2(b). The distribution of photon motion track obtained without intact skull shows a fusiform shape on the whole, and the boundary is not as smooth as that obtained with intact skull. The distribution of photons obtained by removing the skull is fatter than that obtained with intact skull.

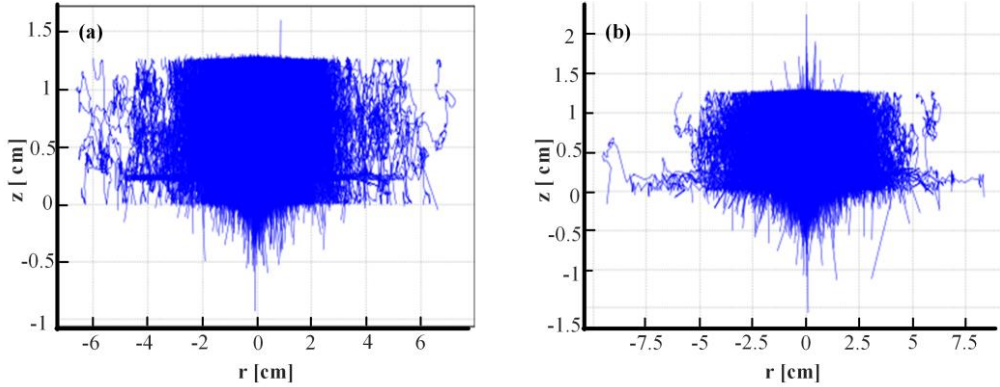

Figure 2. (a) The track diagram of the motion without intact skull. (b) The track diagram of the motion with intact skull.

**3.2 Optical absorption distribution and light intensity distribution of photon propagation in the brain**

Figure 3 shows the distribution of $A_{zr}$ and $F_{zr}$. As can be seen from Figure 3, The optical absorption is mainly concentrated near the central axis. The chance of photons reaching the depths of biological tissues is small. After multiple scattering, the number of photons diverging to the surroundings increased. Under the same conditions, after the skull was removed, the photon transmission appeared deeper, and the lateral direction (that is, the r direction) appeared wider. This is because the saline replaced the scalp and skull. The absorption and scattering effects of the saline were relatively weak, and the photons could get to the depths of the tissue without hindrance, as shown in Figure 3 (a) and Figure 3 (c). There is a clear dividing line on the boundary between the cerebrospinal fluid layer and the brain layer of the tissue. This is because the absorption coefficient and scattering coefficient of the two layers are quite different. In Figure 3(b) and Figure 3(d), due to the strong scattering and absorption of the scalp and skull, the light energy decayed rapidly, and only a few photons could reach the gray matter layer.

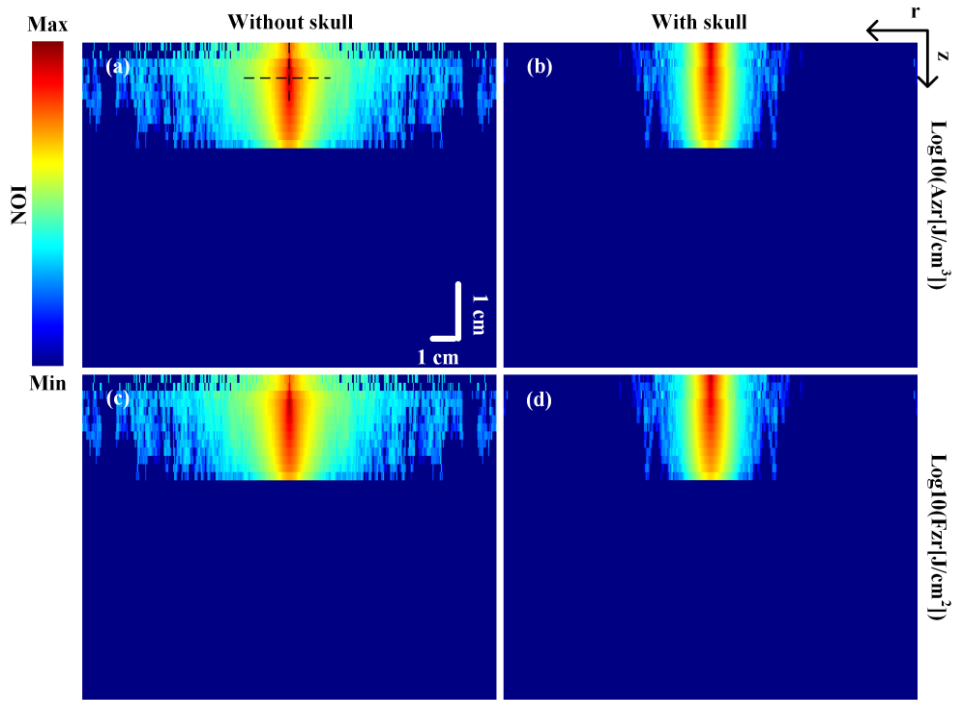

Figure 3. (a) and (b) are the optical absorption density ($A_{zr}$) distribution diagram of photons in mouse brain with intact skull and without intact skull, respectively. (c) and (d) are the fluence rate ($F_{zr}$) distribution diagram of photons in the brain with intact skull and without intact skull, respectively.

## 3.3 Absorption curve of photon propagation in mouse brain

First, the optical absorption density $A_z$ in the depth direction is studied. Figure 4(a) shows that when the skull was not exposed, $A_z$ rapidly dropped sharply from 0.890 cm$^{-1}$ to 0.415 cm$^{-1}$ due to the strong absorption and scattering of the scalp and skull. At this time, the light energy was low, a small amount of photons entered the gray matter. Due to the strong absorption and scattering effects of the gray matter, $A_z$ rose slightly, from 0.415 cm$^{-1}$ to 0.592 cm$^{-1}$, and decreased again. When the depth was 1.35 cm, $A_z$ dropped to 0 cm$^{-1}$. At this time, the photon was basically exhausted and no photon traveled forward. After the skull was removed, due to the weak absorption and scattering effects of normal saline and CSF, $A_z$ was low, which was 0.119 cm$^{-1}$, and dropped slowly. Most of the photons were reserved when they arrived at the grey matter. Due to the strong absorption and scattering effect of the gray matter layer, $A_z$ increased from 0.117 cm$^{-1}$ to 0.812 cm$^{-1}$. When z was about 1.35 cm, $A_z$ decreased 0 cm$^{-1}$. The change of $F_z$ is similar to that of $A_z$, as shown in Figure 4 (b).

In the radial direction, we define the propagation range of photons as the radial distance when $A_r$ dropped to half of the maximum. Figure 4(c) shows that when the photon diffused to a radial distance of about -0.05 cm and 0.05 cm without intact skull, $A_r$ dropped to half, that is, the diffusion width was 0.10 cm. Under the skull, when r was about -0.125 cm and 0.125 cm, $A_r$ dropped to half, that is, the diffusion width was 0.25 cm. It can be seen that, under the same conditions, after the skull was removed, the photon spread more widely in the r direction, which is 2.5 times that of the unopened skull. The change of $F_r$ is similar to that of $A_r$ as shown in Figure 4 (d).

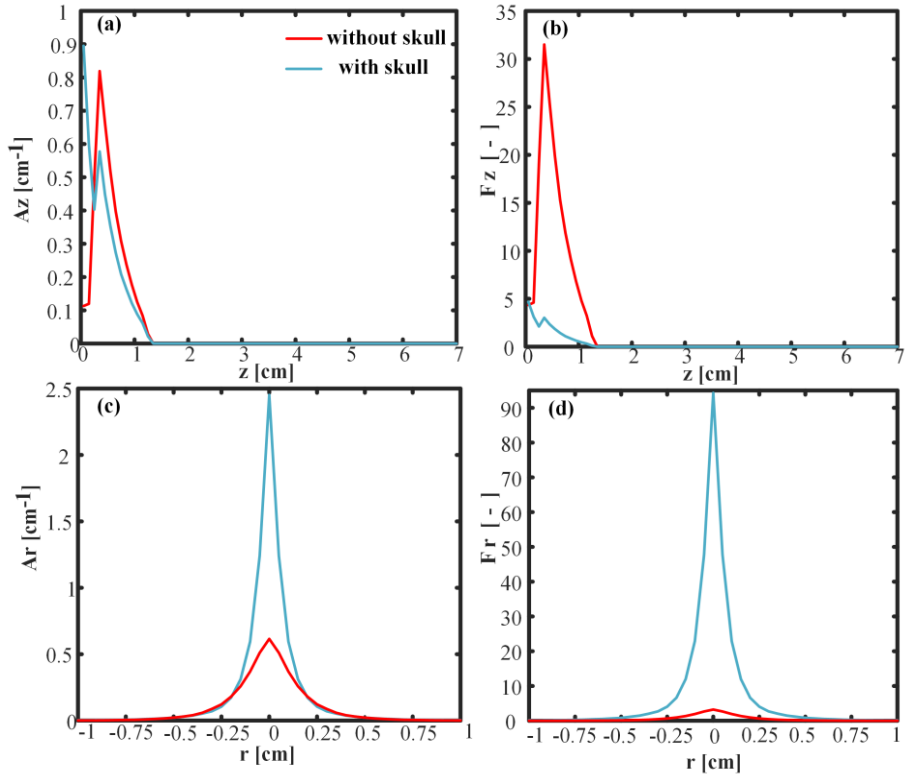

Figure 4. (a) and (b) are the optical absorption density ($A_z$) and fluence rate ($F_z$) of photons in the z direction, respectively. (c) and (d) are the optical absorption density ($A_r$) and fluence rate ($F_r$) of photons in the direction of r, respectively.

## 4. CONCLUSION

We studied the transmission of photons in mouse brain in both with and without intact skull states using Monte Carlo method. Photon track, optical absorption density and fluence rate were used as indicators for analysis. We found that, the photon distribution without intact skull goes farther in both longitudinal and transverse directions compared with that of with intact skull. The distribution of optical absorption density and fluence rate was fusiform and rounder on the whole with intact skull. This study will provide reference and theoretical guidance for the optical imaging of mouse brain and the study of the mouse and human brain.

## REFERENCES

[1] Sofie L. Valk, Felix Hoffstaedter, Julia A. Camilleri, Peter Kochunov, B.T. Thomas Yeo, Simon B. Eickhoff., "Personality and Local brain Structure: Their Shared genetic basis and copyright," NeuroImage, 220, 117067 (2020).
[2] Shuaib A., Butcher K., Mohammad A. A., et al. "Collateral blood vessels in acuteischaemic stroke: a potential therapeutic target," Lancet Neurology, 10(10), 909-921 (2011)
[3] Yan D., Zhang Z., Luo Q., et al. "A Novel Mouse Method Based on Dynamic Contrast Enhanced micro-CT Images," Plos One, 12(1), E0169424 (2017)
[4] Menon U, Gentry-Maharaj A., Hallett R., et al. "Sensitivity and specificity of Multimodal and ultrasound screening for ovarian cancer, and stage distribution of detected programmers," Uniform Screening of the UK Collaborative Trial of Ovarian Cancer Screening (UKCTOCS). Lancet Oncol, 10(4), 327-340 (2009).
[5] Cai W., Chen K., Li Z. B., et al. "Dual-function probe for PET and near infrared imaging of tumor Vasculature," Journal of Nuclear Medicine, 48(11), 1862-1870 (2007).
[6] Owen R. S., Carpenter J. P., Baum R. A., et al. "Magnetic resonance Imaging of graphically attended in Peripheral arterial occlusive disease," New England Journal of Medicine, 326 (24), 1577-1581 (1992).
[7] Povazay B., Bizheva K., Hermann B., et al. "Enhanced visualization of choroidal Using ultrahigh Resolution ophthalmic OCT at 1050 nm," Optics Express, 11(17), 1980-1986 (2003).
[8] Cheung, a. T., Chen, p.c., Larkin e. C., et al. "The Microvascular abnormalities in sickle cell diseases: A computer – assisted intravital microscopy study," Blood, 99(11), 3999-4005 (2002)
[9] Mehrabian H., Lindvere L., Stefanovic B., et al. "A constrained independent component analysis technique for artery–vein separation of two-photon laser scanning microscopy images of the cerebral microvasculature," Medical Image Analysis, 16(1), 239-251 (2012).
[10] Momose and Okada E. "Frequency Domain Monte Carlo prediction of light propagation in bifocal bodies with clear Region," Proc. SPIE, 3566, 28-36 (1998)